\documentclass[conference]{IEEEtran}
\IEEEoverridecommandlockouts

\usepackage{comment}
\usepackage{cite}
\usepackage{tipa}
\usepackage{makecell}
\usepackage{amsmath,amssymb,amsfonts}
\usepackage{algorithmic}
\usepackage{graphicx}
\usepackage{textcomp}
\usepackage{xcolor}
\usepackage{tipa}
\def\BibTeX{{\rm B\kern-.05em{\sc i\kern-.025em b}\kern-.08em
    T\kern-.1667em\lower.7ex\hbox{E}\kern-.125emX}}
\begin{document}

\makeatletter
\twocolumn[%
\begin{@twocolumnfalse}%
\vspace{-0.5em}
\begin{center}
© 20XX IEEE.  Personal use of this material is permitted.  Permission from IEEE must be obtained for all other uses, in any current or future media, including reprinting/republishing this material for advertising or promotional purposes, creating new collective works, for resale or redistribution to servers or lists, or reuse of any copyrighted component of this work in other works.
\end{center}
\vspace{1em}%
\end{@twocolumnfalse}%
]
\makeatother

\title{ Acoustic to Articulatory Speech Inversion for Children with Velopharyngeal Insufficiency\\
{\footnotesize \textsuperscript{}}
}
\title{Acoustic to Articulatory Speech Inversion for Children with Velopharyngeal Insufficiency\\
{\footnotesize \textsuperscript{}}
\thanks{This work was supported by NSF Grant No. BCS2141413.}
}
\author{
    \IEEEauthorblockN{
        Saba Tabatabaee\IEEEauthorrefmark{1},
        Suzanne Boyce\IEEEauthorrefmark{2},
        Liran Oren\IEEEauthorrefmark{3},
        Mark Tiede\IEEEauthorrefmark{4},
        Carol Espy-Wilson\IEEEauthorrefmark{1}
    }
    \IEEEauthorblockA{\IEEEauthorrefmark{1}Department of Electrical and Computer Engineering, University of Maryland, College Park, USA}
    
    \IEEEauthorblockA{\IEEEauthorrefmark{2}Department of Communication Sciences and Disorders, University of Cincinnati, USA}
    
    \IEEEauthorblockA{\IEEEauthorrefmark{3}Department of Otolaryngology–Head and Neck Surgery, University of Cincinnati, USA}
    
    \IEEEauthorblockA{\IEEEauthorrefmark{4}Department of Psychiatry, Yale University, USA}
    
}

\maketitle

\begin{abstract}
Traditional clinical approaches for assessing nasality, such as nasopharyngoscopy and nasometry, involve unpleasant experiences and are problematic for children. Speech Inversion (SI), a noninvasive technique, offers a promising alternative for estimating articulatory movement without the need for physical instrumentation. In this study, an SI system trained on nasalance data from healthy adults is augmented with source information from electroglottography and acoustically derived F0, periodic and aperiodic energy estimates as proxies for glottal control. This model achieves 16.92\% relative improvement in Pearson Product-Moment Correlation (PPMC) compared to a previous SI system for nasalance estimation. To adapt the SI system for nasalance estimation in children with Velopharyngeal Insufficiency (VPI), the model initially trained on adult speech was fine-tuned using children with VPI data, yielding an 7.90\% relative improvement in PPMC compared to its performance before fine-tuning.
\end{abstract}

\section{Introduction}

Nasality is one of the key contrastive features in speech production \cite{fry2004phonics}. Nasal consonants and nasalized vowels are widely present across the world’s languages. Nasality is regulated by how much the velopharyngeal (VP) valve is constricted, which affects both airflow and the acoustic coupling between oral and nasal cavities. This process involves the coordinated movement of the velum and pharyngeal walls. During normal speech, the VP valve operates over a range between fully open, for nasal sounds, and fully closed, for occlusive sounds that require oral pressure buildup.  The VP valve may be open or partially open for oral vowels and semivowels, with patterns determined by constriction location, dialect and language \cite{bell1995anticipatory}.  

For speakers whose VP valve does not close properly due to VP dysfunction, the normal speech patterns are disrupted. The term velopharyngeal insufficiency (VPI) describes VP dysfunction due to abnormal structure. This dysfunction causes speech disorders because the VP valve does not separate the oral and nasal cavities during normal oral speech. Speech disorders due to VPI are marked by an undesired balance of resonance and/or airflow in the nasal cavity during speech. 

"Resonance" disorders include too much sound energy in the nasal cavity on oral sounds (hypernasality) or too little sound energy in the nasal cavity on nasal sounds (hyponasality). VPI also results in airflow leaking into the nasal cavity during non-nasal phonemes, referred to as nasal emission, which can be inaudible (typically with hypernasality) or audible. The severe form of audible nasal emission is called nasal turbulence (or nasal rustle). Hypernasality is the most common disorder in patients with VPI and is always associated with voiced, rather than voiceless, speech sounds. Audible nasal emission is the second most defining feature of children with VPI and is most noticeable on plosive consonants and fricatives.    

Assessment of VPI starts with a perceptual evaluation by an experienced speech-language pathologist. Perceptual assessment is necessary because treatment is only indicated when a problem with speech is perceived. Once a patient’s speech has been determined to show the effects of VPI, the next step typically includes a form of instrumental measurement that attempts to identify its etiology.

Generally, the procedures that assess VPI further can be divided into two broad categories: direct and indirect measurement techniques. Direct measurements can visualize VP structures and motion and provide information such as the opening size and location, closing pattern, and timing synchronization with other articulators. Although there is no ideal direct measurement technique, videofluoroscopy and nasopharyngoscopy are the most commonly used procedures. Both techniques are considered limited because they capture specific planes and reduce the 3D anatomy of the VP valve to a 2D image, making relevant anatomical information harder to interpret \cite{watterson2021reliability}. The high cost of instrumentation, as well as difficulties in imaging children, is the main reason why not all clinicians, even those specializing in resonance disorders, have access to direct measurement of VP function. 

Indirect measurement techniques provide information from which VPI can be inferred. Nasometry is an indirect measure that provides objective data regarding the function of the VP valve. As such, it is probably the most common technique clinicians use \cite{kummer2018cleft}.  The technique provides a nasalance score, a measure that can be compared with standardized norms. High nasalance scores on non-nasal sounds indicate VPI. Nasometry and similar indirect measures are favored for use in the clinic due to their simplicity, low cost and non-invasive nature, especially with children.  Clinicians also use nasalance scores to monitor VPI treatment progress, with a reduction in nasalance score to indicate VPI improvement.

In healthy adults, a strong correlation has been demonstrated between nasalance and the degree of VP constriction for nasal targets, and a moderate to good correlation for full audio recordings, as validated by High-Speed Nasopharyngoscopy (HSN) images \cite{siriwardena2024speaker}. Recent advances in deep learning have led to the development of Speech Inversion (SI) systems as a noninvasive and cost-effective method for estimating nasalance directly from audio signals \cite{siriwardena2024speaker,feng2024wav2nas,siriwardena2023speaker}. In contrast to traditional techniques such as nasometry and nasopharyngoscopy, which require specialized equipment, SI systems eliminate the need for physical instrumentation, offering a more accessible and affordable tool for both clinical and remote settings. Previous research has used the  audio spectrograms and Mel-Frequency Cepstral Coefficients (MFCCs) as features extracted from speech in SI tasks \cite{siriwardena2023speaker, siriwardena2022acoustic}. However, recent progress in Self-Supervised Learning (SSL) has shifted the development of SI systems toward more advanced and robust feature extraction methods, surpassing the traditional acoustic features\cite{siriwardena2024speaker, cho2023evidence, attia2024improving,tabatabaee25b_interspeech}. 

A study by Siriwardena et al. \cite{siriwardena2024speaker} demonstrated that a BiGRNN model leveraging HuBERT-Large SSL representations significantly outperformed a Temporal Convolutional Network (TCN) that relied on spectrogram based feature extraction for nasalance estimation. Moreover, the study showed that the SSL-based SI system demonstrated better performance across different speech corpora, highlighting the robustness and generalization of SSL pre-trained models to unseen corpora. In a study by Cho et al. \cite{cho2023evidence} it was shown that WavLM-Large surpassed SSL models, such as HuBERT-Large, Wav2Vec 2.0 and TERA, as well as traditional acoustic features in SI tasks. Building upon these findings, this study leverages SSL representations, including HuBERT-Large, Wav2Vec 2.0-Large, and WavLM-Large, to capture richer speech features, thereby enhancing the overall performance of the SI system.

Several studies \cite{siriwardena2023secret, siriwardena2023speaker} have shown that incorporating Source Features (SFs) such as the aperiodic energy (non-periodicity), periodic energy (periodicity), and fundamental frequency (F0) extracted from speech signals into SI systems improves their performance. A study by Siriwardena et al. \cite{siriwardena2023speaker} demonstrated that the combination of the envelope of electroglottography (EGG) signal and SFs enhances nasalance estimation, with the EGG envelope serving as an additional proxy for glottal activity.   However, the individual contributions of these glottal features to nasalance estimation were not investigated. Building on these prior works, the present study integrates SFs and the envelope of the EGG signal into the SI framework to improve the estimation of nasalance. To gain clearer insight into their respective roles, in the present study an ablation analysis is performed to evaluate the individual impact of the EGG envelope and SFs on nasalance estimation.

The limited availability of speech data from children has led investigators to study whether “fine-tuning” systems developed using adult speech with a smaller amount of children’s speech data, can produce effective results with children’s speech in tasks such as automatic speech recognition, speaker verification, and other speech related applications \cite{tabatabaee25_interspeech,rolland2024introduction}. Prior studies \cite{benway2023acoustic,benway2025subtyping, benway2024examining,benway2025perceptual} have demonstrated that SI systems trained on adult speech can be effectively applied to children's speech. However, these studies did not explore the potential benefits of fine-tuning these systems with children’s speech data to further improve their performance. Therefore, this study investigates the effect of fine-tuning for SI tasks by first proposing an SI system for nasalance estimation in healthy adults, and then adapting the model for children with VPI through fine-tuning.

A key challenge in developing a SI system to track VP constriction for children with VPI is the lack of available fine-grained ground truth nasalance data showing the time course of VP movement. To address this gap, we start with a baseline of an SI system initially trained on adult speech and subsequently apply fine-tuning using a small dataset of speech from children with VPI.  This two stage approach allows for the effective adaptation of the nasalance estimation system to the pediatric VPI children. To the best of our knowledge, this is the first study to develop a SI system for children with VPI that estimates nasalance directly from an audio speech signal.

\textbf{The key contributions} of this work are as follows:
\begin{itemize}
\item Enhanced estimation of nasalance for adult speech using a proposed multi-task learning SI system, demonstrating improved performance over the approach in \cite{siriwardena2024speaker}.
\item A comparative evaluation of the adult SI systems using three SSL representation models of HuBERT-Large, wav2vec 2.0-Large, and WavLM-Large.
\item An ablation study analyzing the impact of incorporating the envelope of EGG and SFs on the nasalance estimation relative to the ground truth. 
\item Introduction of a novel SI system capable of estimating nasalance in children with speech disorders due to VPI.
\end{itemize}

\section{Dataset Description}
\subsection {Healthy-Adult dataset}
Twenty-four adult speakers, assessed by a speech-language pathologist to have normal speech were selected. Compared to \cite{siriwardena2024speaker}, we collected data from four additional subjects. Participants included 20 native English speakers, 3 native French speakers, and 1 native Sinhala speaker. The participants were asked to read sections from well-known passages, including the "Grandfather Passage" \cite{darley1975motor}, Harvard Sentences \cite{rothauser1969ieee},  as well as passages from \cite{krakow1989articulatory} and \cite{westbury1994speech}. The dataset comprised a total of 993 utterances and 2.04 hours of speech. 

\subsection {VPI-Child dataset}
Fourteen children diagnosed with VPI were selected from patients at the VPI Clinic at Cincinnati Children’s Hospital Medical Center. As part of the routine clinical assessment, a nasopharyngoscopy exam was performed jointly by an otolaryngologist and a speech-language pathologist to assist in the diagnosis of children with VPI. All participants were native English speakers, consisting of 5 females and 9 males, ranging in age from 3 to 15 years, with a mean age of 7.9 years. A high proportion of the children showed hypernasality (9 out of 14), audible nasal emission (6 out of 14), and nasal rustle (3 out of 14), with some children having an overlapping combination of these features. The recorded sentence set included phonemes targeting key articulatory features such as alveolar, bilabial, nasal, sibilant, and velar sounds. Children’s diagnoses ranged from mild to moderate hypernasality, with some children also exhibiting variable instances of audible nasal emission/nasal rustle. The resulting dataset contained 318 utterances, with a total duration of 7.04 minutes. 

\section{Methodology}
\subsection {Nasometry setup and data preparation}
In this study, we developed a nasometry system inspired from \cite{siriwardena2024speaker} for collecting data from healthy adults and children diagnosed with VPI in a separate session.

As shown in Figure 1, oral and nasal signals were captured using separate microphones mounted at the center of a 5mm thick aluminum partition plate. The plate was placed against the subject’s upper lip, creating an acoustic barrier that separated the oral and nasal audio recordings. Windscreens were placed over the microphones to shield them from interference caused by airflow directed at them. Oral and nasal signals were recorded at a sampling rate of 51.2 kHz. As illustrated in Figure 1, HSN images were captured using a flexible endoscope connected to a high speed video camera. The recordings were made at a resolution of 304 × 256 pixels and a frame rate of 1000 frames per second. The HSN images were utilized to assess the quality of nasalance ground truth data obtained from the nasometry setup. 

The ground truth nasalance values were derived from simultaneously recorded oral and nasal microphone signals. To eliminate baseline wander and low-frequency noise, the signals from the Healthy-Adult dataset were high-pass filtered at 20 Hz, and those from the VPI-Child dataset were high-pass filtered at 10 Hz. The acoustic energy was subsequently calculated as the root mean square of the signals and smoothed using a 25 ms rectangular moving average filter. The nasalance measure was calculated by utilizing the nasal acoustic energy (AEnasal) and oral acoustic energy (AEoral), as defined in equation \ref{eq:NAS}:
\begin{equation}
\text{Nasalance} = \frac{AEnasal}{AEnasal + AEoral}
\label{eq:NAS}
\end{equation}
The nasalance was resampled to 100 Hz and normalized across all utterances within each dataset to a range of -1 to 1. 

In addition to the collection of nasal and oral signals, the EGG signal was recorded simultaneously for adults using electrodes placed on the neck at the level of the thyroid prominence. The EGG setup used for the Healthy-Adult dataset was not employed for the VPI-Child dataset. The ground truth for the EGG envelope (EGG-env) parameter was derived from the EGG signal, which was recorded at a sampling rate of 51.2 kHz. The EGG signal was high-pass filtered using a 20 Hz cutoff frequency, and its envelope was extracted by computing the magnitude of the Hilbert transform. This envelope was then resampled to 100 Hz and normalized across all utterances within the Healthy-Adult dataset to a range of -1 to 1.

\begin{figure}[htbp]
    \hfill
    \centering
    \vspace{-1mm}
     \includegraphics[width=0.49\textwidth, height=0.17\textheight]{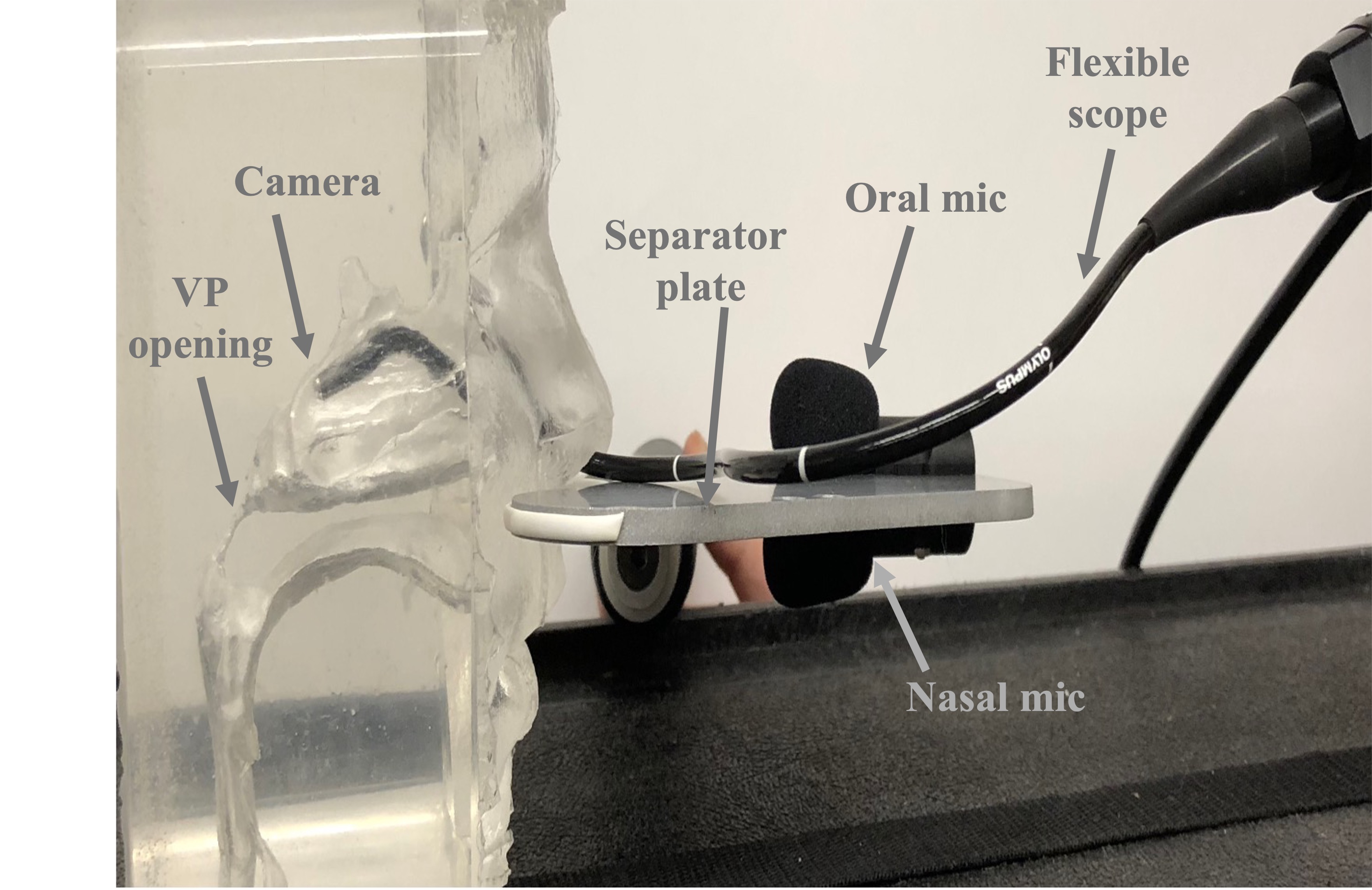}  
    \caption{Illustration of the nasometry experimental setup. Oral and nasal microphones are mounted on a separator plate. HSN measurements are captured using a flexible scope connected to a high speed video camera.
    }
    \label{fig:nasal_si}
   
\end{figure}

To split the Healthy-Adult and VPI-Child data into train, development and test set, a speaker-independent approach was used to ensure that the SI model's performance was evaluated on unseen speakers, thereby enhancing generalization and reducing speaker-specific bias. For the Healthy-Adult dataset, the same test set used in \cite{siriwardena2024speaker} was used in this study to enable a fair and consistent comparison with prior work. A detailed summary of the Healthy-Adult dataset is presented in Table I. 

Due to the limited size of the VPI-Child dataset, a 5-fold cross-validation strategy was implemented to ensure reliable evaluation. In each fold, data from 8 participants were used for training, 3 for development, and 3 for testing.

\begin{table}[ht!]
\caption{Healthy-Adult dataset description.} 
\centering 
\label{speaker_verification_models}
\resizebox{\columnwidth}{!}{
    \begin{tabular}{|l|c|c|c|c|}
        \hline
        Split & \# Subjects & Language & Duration (hour) & \# Utterance \\
        \hline
        Train & 20 & \makecell{ English: 16 \\French: 3 \\  Sinhala: 1} & 1.70 & 858\\
        \hline
        Development & 2 & English & 0.18 & 65\\
        \hline
        Test & 2 & English &  0.16 & 70\\    
        \hline
    \end{tabular}
       }
\end{table}
\begin{table*}[htbp]
\caption{PPMC scores of the SI systems using various SSL models on the Healthy-Adult test set.}
\centering
\label{speaker_verification_models}
 \begin{tabular}{|l|c|c|c|c|c|c|c|}
        \hline
        SI System & Training Data &SSL Model& VP & EGG-env & Per & Aper & F0 \\
        \hline
        Baseline \cite{siriwardena2024speaker} & Data from  \cite{siriwardena2024speaker} & HuBERT-Large & 0.8115 &0.8330 & 0.8373 & 0.8542 & 0.8385 \\
        \hline
         Baseline \cite{siriwardena2024speaker}  & Current study  & HuBERT-Large  &0.8904 & 0.8331 & 0.7320 & 0.5348 & 0.7669 \\

        \hline
        SI & Current study &Wav2vec 2.0-Large & 0.9409& 0.8535& 0.9381 & 0.8186 & 0.8044 \\
        \hline
     SI& Current study &HuBERT-Large & 0.9437& 0.8529 & 0.9376& 0.8201 &  0.7983\\
                \hline
         SI & Current study &WavLM-Large &0.9488 & 0.8646 & 0.9415 &  0.8321 & 0.8062 \\
        \hline
    \end{tabular}
\end{table*}
\subsection {Developing SI system}
Audio recorded from both the oral and nasal microphones was combined into a single signal for developing the SI system. This signal was then down sampled to 16 kHz to match the sampling rate required by the SSL feature extractors. Unlike the approach in \cite{siriwardena2024speaker}, which used fixed two-second segments for training, we employed randomly selected variable-length segments ranging from 2 to 5 seconds. This approach was designed to improve the robustness of the SI system to variations in audio duration. 

The SI system was developed using a multi-task learning approach, where the first task estimates the nasalance measure, and the second task estimates the EGG-env and three SFs, which are aperiodicity (Ap), periodicity (Per), and F0 as illustrated in Figure 2.  In this study, the nasalance estimated by the SI system is referred to as the VP Tract Variable (TV). The ground truth values for the three SFs were extracted from speech signals using the APP detector \cite{deshmukh2005use}.

The SI system was developed using various SSL pre-trained models, including HuBERT-Large \cite{hsu2021hubert}, Wav2Vec 2.0-Large \cite{baevski2020wav2vec}, and WavLM-Large \cite{chen2022wavlm}. These three embeddings were compared, and the best performing SSL model in the development of the SI system was proposed as the optimal feature extractor. Representations from all 25 hidden layers of each SSL model were stacked, and a two-dimensional convolutional layer was used to compute a weighted sum of these representations, yielding a single-layer embedding. This embedding was then passed through two bidirectional Gated Recurrent Unit (GRU) layers (each with 256 units and a dropout rate of 0.3), followed by a dense layer with 128 hidden units. The output was subsequently upsampled by a factor of two to align with the target sampling rate of 100 Hz required for VP TV, EGG-env, and the three SFs, as the original output from the SSL models had a sampling rate of 50 Hz. This upsampling step was followed by batch normalization and a dropout layer with a rate of 0.3.

The SI system was optimized using the ADAM optimizer with a learning rate of 5e-4 and a batch size of 8, both selected through the grid search. To promote efficient convergence and mitigate overfitting, a learning rate scheduler (ReduceLROnPlateau) was employed in conjunction with early stopping, with a patience of five epochs. In our multi-task learning framework, we adopt the same loss formulation for both tasks, defined as a weighted combination of Pearson Correlation (PC) and Root Mean Square Error (RMSE), as shown in Equation~\ref{eq:loss_function}. The total loss is then computed as the sum of the individual losses obtained for each task.

\begin{equation}
Loss  = (1 - \text{PC}) + (\alpha) \cdot \text{RMSE} 
\label{eq:loss_function}
\end{equation}

In equation 1, \(\alpha\) was set to 0.2. This value was empirically determined by evaluating multiple \(\alpha\) values and selecting the one that yielded the best performance.

\begin{figure}[htbp]
    \hfill
    \vspace{-1mm}
     \includegraphics[width=0.49\textwidth, height=0.2\textheight]{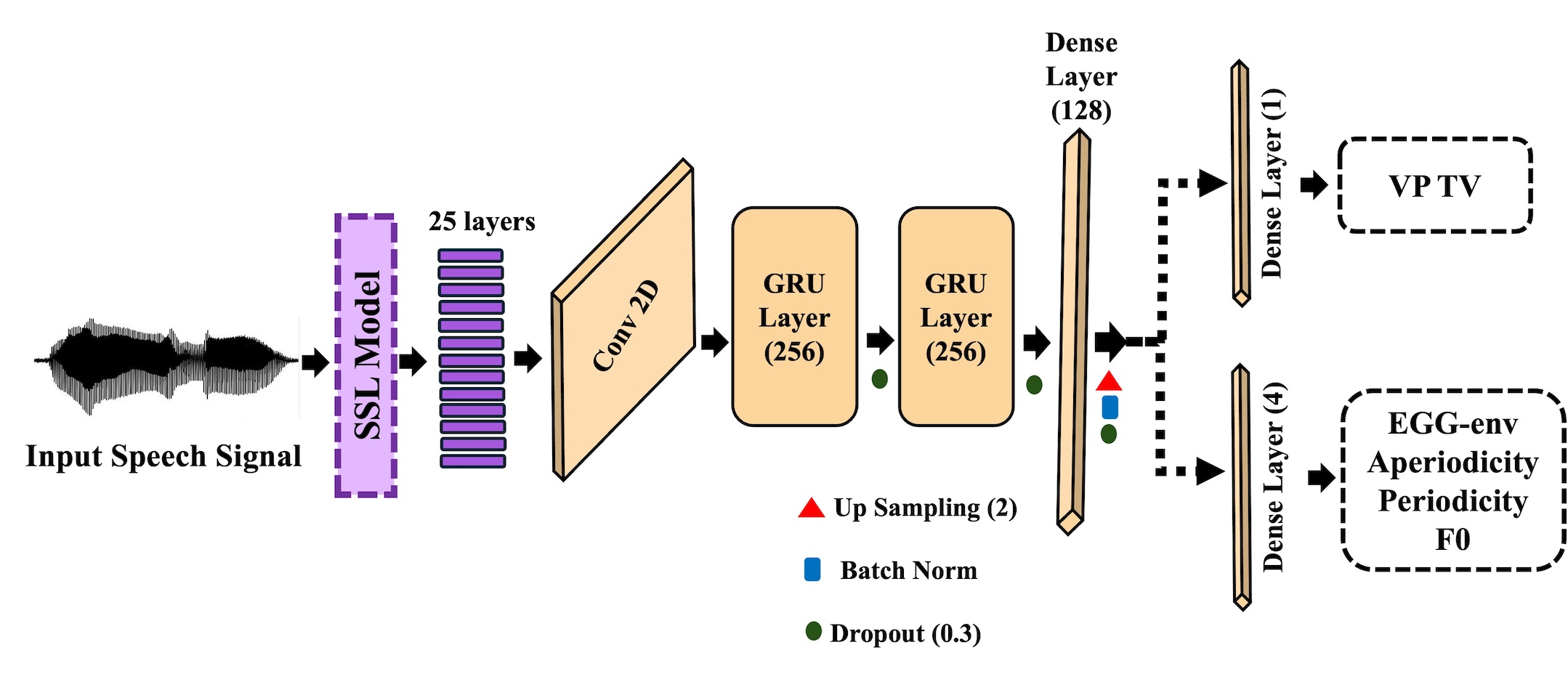}  
     \vspace{-5mm}
    \caption{Proposed model architecture for SI system.}
    \vspace{-5mm}
    \label{fig:nasal_si}
   
\end{figure}

\subsubsection {Fine-tuning SI system using VPI-Child dataset} 

As noted above, the SI system was initially trained on the Healthy-Adult speech dataset. Since the VPI-Child dataset does not contain EGG signals, we used a version of the SI system that excludes EGG-env features, enabling us to fine-tune the SI system using the VPI-Child dataset. This SI system is capable of estimating the VP TV and three SFs, and was later used to adopt the SI systems to VPI child speech. To address the limited size of the VPI-Child dataset, we employed a fine-tuning strategy rather than training the system from scratch.

\section{Results and Discussion}
\subsection{Results for the SI system trained on the Healthy-Adult dataset}
The SI system architecture proposed in \cite{siriwardena2024speaker}, which utilizes the HuBERT-Large pre-trained model, was used as the baseline for performance evaluation of the proposed SI system. To ensure a fair comparison, the evaluation was conducted using the identical test set as in \cite{siriwardena2024speaker}. The first row in Table II presents the original results reported in \cite{siriwardena2024speaker}, while the second row shows the performance of their model architecture retrained on our dataset, which includes the original data from \cite{siriwardena2024speaker} along with speech from four additional subjects collected in this study. Both systems (the proposed multi-task model and the retrained baseline) were trained and evaluated on the same dataset to ensure a fair comparison.

The results in Table II show that the proposed multi-task SI architecture and training strategy (employing variable-length segments of 2 to 5 seconds instead of the fixed 2-second segments used in \cite{siriwardena2024speaker}) consistently outperform the baseline SI model in VP TV estimation. This performance advantage remains consistent both when the baseline model is trained on the original dataset from \cite{siriwardena2024speaker} and when trained on the extended dataset. These findings highlight the effectiveness of our proposed training strategy and SI model architecture in enhancing the development of SI systems for VP TV estimation.

Moreover, the proposed multi-task SI architecture using the WavLM-Large pre-trained model achieved the best performance in estimating the VP TV parameter, surpassing both the baseline and other SSL-based SI systems. Specifically, the SI system with WavLM-Large embeddings achieved a 6.56\% relative improvement in the PPMC score for VP TV parameter estimation compared to the baseline SI system trained on our dataset, and a 16.92\% relative improvement compared to the baseline model trained on the dataset from \cite{siriwardena2024speaker}.

\subsection{Ablation study for the SI system}

Table III presents an ablation study on the  best performing SI system, which incorporates the WavLM-Large SSL pre-trained model, to evaluate the impact of the acoustically-derived  three SFs and the EGG-env on VP TV parameter estimation. In the first row, both EGG-env and the three SFs were excluded, creating a single-task setup where only the VP TV was estimated. In subsequent configurations, either the EGG-env or the three SFs were included alongside the VP TV, allowing for a multi-task learning approach. In both cases, the inclusion of glottal-related features led to improved PPMC scores for VP TV estimation, demonstrating that integrating complementary glottal information into the multi-task learning framework enhances the accuracy of VP TV estimation. 

Finally, the best performance in VP TV estimation was achieved when VP TV was estimated together with EGG-env and 3 SFs, consistent with the work in \cite{siriwardena2023speaker}. These findings highlight the fine grained interplay between vocal fold dynamics and VP movement, and demonstrate that enriching the SI system with glottal features enhances VP TV estimation.

\begin{table}[htbp]
\caption{Ablation study of the SI system. The table reports the PPMC scores on the Healthy-Adult test set.}
\vspace{-1mm} 
\centering 
\label{ablation}
    \begin{tabular}{|l@{\hskip 2pt}|c@{\hskip 4pt}|c@{\hskip 4pt}|c@{\hskip 3pt}|c@{\hskip 3pt}|c@{\hskip 3pt}|}
        \hline
        Excluded Param & VP & EGG-env  &Per& Aper& F0  \\
        \hline
     EGG-env , 3 SFs  & 0.9444 &-& - & - & - \\
     \hline
        EGG-env    & 0.9466&- & 0.9465
 &0.8377& 0.8197\\
             \hline
    3 SFs& 0.9470& 0.8593 & - & - & - \\
     \hline
-  &0.9488 & 0.8646 & 0.9415 &  0.8321 & 0.8062 \\
             \hline             
    \end{tabular}

\end{table}

\begin{table*}[htbp]
\caption{Average PPMC scores of the SI system on the VPI-Child test set across five folds, reported before and after fine-tuning. Values in parentheses indicate the standard deviation.}
\centering
\label{speaker_verification_models}
\begin{tabular}{|l|c|c|c|c|c|c|c|}
    \hline
    System &  Pre-training Data & Fine-tuning Data & SSL Model & VP & Per & Aper & F0 \\
    \hline
   SI           & Healthy-Adult     & --       &WavLM-Large  &0.6357 (0.14) & 0.8826 (0.04) & 0.8431 (0.01) & 0.5588 (0.07) \\
   \hline
    SI-FT      & Healthy-Adult      & VPI-Child  & WavLM-Large  & 0.6859 (0.08)  & 0.8765 (0.04) & 0.8384 (0.01) & 0.6215 (0.06) \\
\hline

\end{tabular}
\end{table*} 

\subsection{Cross-corpus evaluation of the proposed SI system }
To further assess the generalization of the proposed SI system, we analyzed its performance for the utterance “Say amp again,” drawn from a different dataset that included simultaneously recorded nasal, oral, and far-field signals. Nasalance was computed using the oral and nasal signals, following the definition provided in Equation 1. Far-field audio was used to obtain VP TV using the SI system.

As shown in Figure 3, the SI system  accurately estimates the articulatory constriction patterns for each nasal consonant, demonstrating a close correspondence with the ground truth. Notably, for the instances of /m/ in “amp” and /n/ in “again,” the VP TV estimated by the SI system exhibits two peaks that align with the nasalance ground truth. These results suggest that the proposed SI system generalizes well to speech data from a corpus not encountered during training.

\begin{figure}[htbp]
    \hfill
    \centering
    \vspace{-1mm}
     \includegraphics[width=0.5\textwidth, height=0.32\textheight]{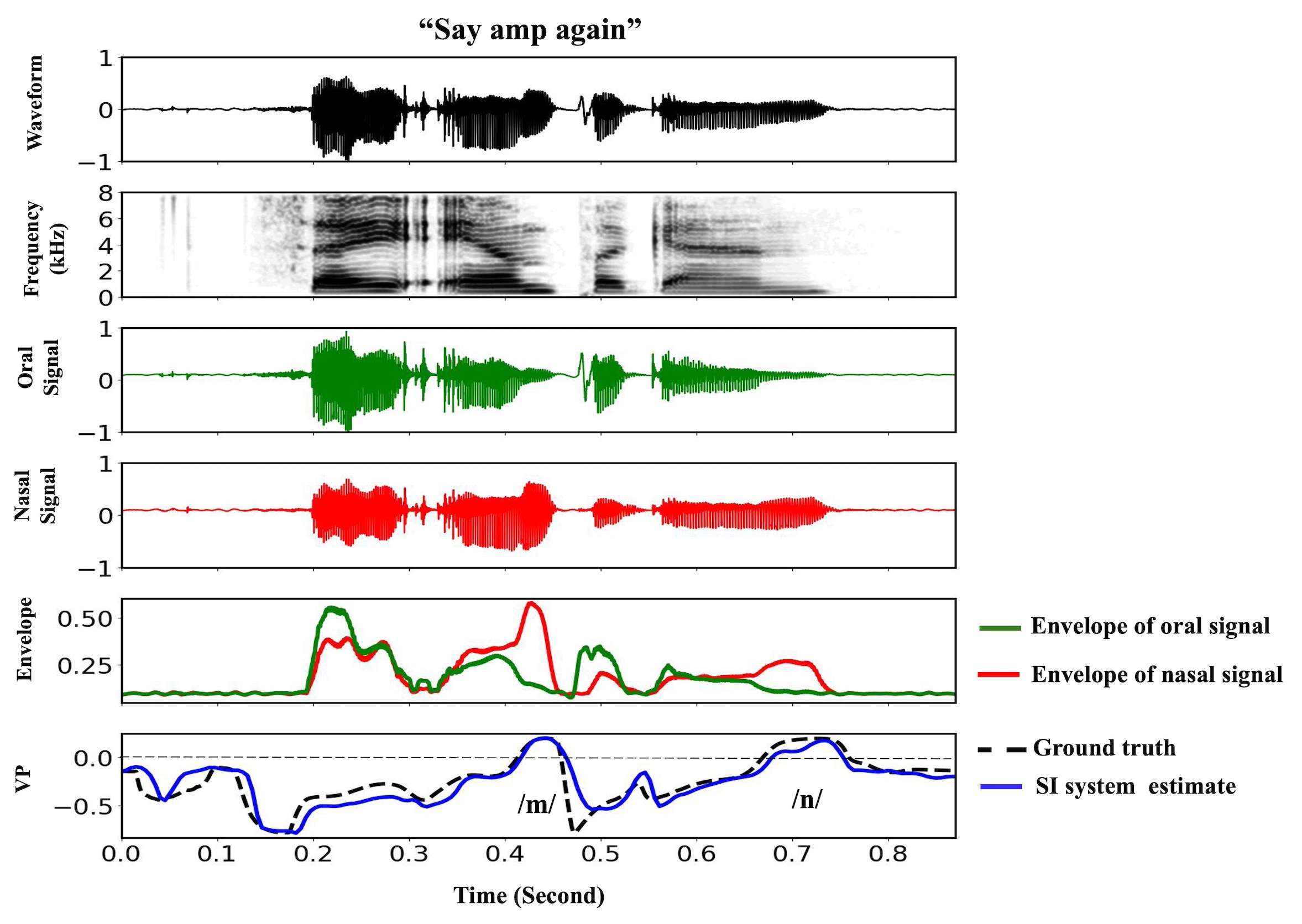}  
    \caption{Waveforms, envelope of oral and nasal signals and spectrogram are presented, followed by a comparison between the ground truth nasalance and the corresponding estimates produced by the SI system for the utterance “Say amp again”. The second row displays the spectrogram from the far-field signal illustrating the signal’s frequency content over time, with color intensity indicating the strength of each frequency component. The third and fourth rows show the near-field oral and nasal signals, respectively. The fifth row displays the smoothed signal envelopes, with the oral signal shown in green and the nasal signal in red. The sixth row shows the estimated VP TV from the SI system alongside the ground truth nasalance. 
    }
    \label{fig:nasal_si} 
\end{figure}

\subsection{The SI systems performance on the VPI-Child dataset before and after fine-tuning}

Table IV presents the average evaluation results of the SI system across five folds of the VPI-Child test set, both before and after fine-tuning. As shown in Table IV, fine-tuning with the VPI-Child dataset led to performance improvement for estimating VP TV, despite the limited amount of in-domain data. The SI system achieved a relative improvement of 7.90\% in VP TV estimation after fine-tuning. This result underscores the importance of domain adaptation in SI tasks, demonstrating that even small amounts of disorder-specific child speech can significantly enhance SI systems performance. 
To provide a more comprehensive evaluation of the two-step training approach (pre-training on healthy adult speech followed by fine-tuning on the VPI-Child dataset), we trained the SI system from scratch exclusively on the VPI-Child dataset, achieving a PPMC score of 0.6388 in VP TV estimation. Therefore, compared to training from scratch using only the VPI-Child dataset, the two-step training strategy yielded a relative improvement of 7.37\%, highlighting the effectiveness of fine-tuning in limited in-domain data scenarios.


\subsection{Nasal emissions in VPI speech disorder}
Figure 4 illustrates an example of a nasal rustle sound occurring during a stop consonant in a child diagnosed with VPI. A comparison between the estimated VP TV for the utterance “Take a turtle” between the SI systems before and after fine-tuning on VPI child data is shown.

The fine-tuned system (SI-FT) demonstrates an enhanced ability to accurately track the ground truth nasalance.  As the example shows, the child with VPI was not able to completely close the VP valve, leading to some audible nasal emission, or nasal rustle as well as nasal resonance.  In Figure 4, both the ground truth nasalance and the estimated nasalance from the SI-FT system exhibit two peaks, one corresponding to the /t/ in “take” and another to the first /t/ in “turtle” which align with audible nasal emissions produced by the child with VPI.

\begin{figure}[htbp]
    \hfill
    \centering
    \vspace{-1mm}
     \includegraphics[width=0.5\textwidth, height=0.32\textheight]{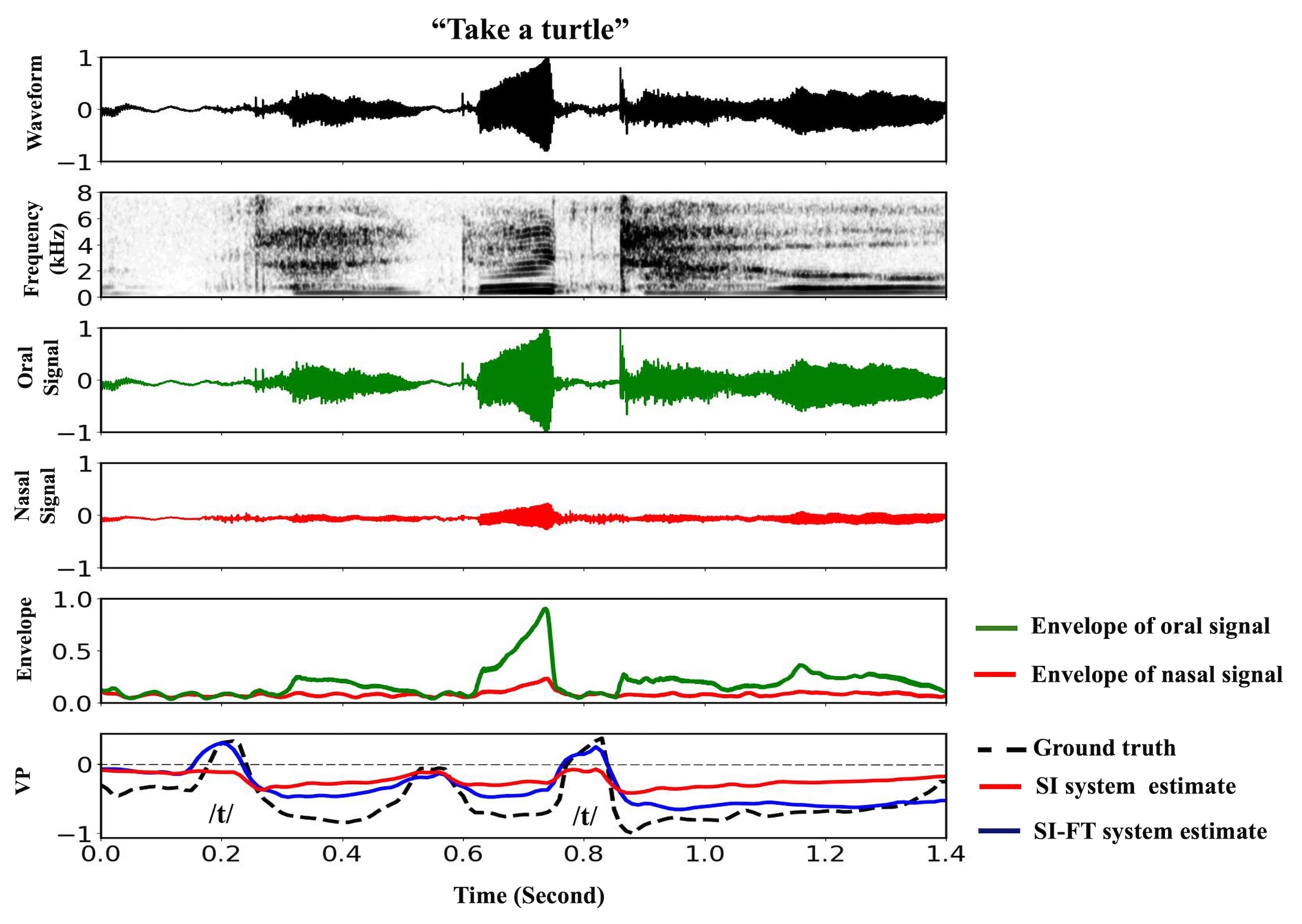}  
    \caption{
    Waveforms, envelope of oral and nasal signals and spectrogram of the utterance “Take a turtle” from a child subject are presented. A comparison between the ground truth nasalance and the corresponding estimates produced by the SI system before and after fine-tuning are shown. The second row displays the spectrogram from the combined signal. The third and fourth rows show the oral and nasal signals, respectively. The fifth row displays the smoothed signal envelopes, with the oral signal shown in green and the nasal signal in red. The sixth row displays the estimated nasalance from the SI and SI-FT systems alongside the ground truth nasalance.
    }
    \label{fig:nasal_si}
   
\end{figure}

Figure 5 presents an example of vowel nasalization observed in the utterance “Suzie sees”, as produced by a pediatric child diagnosed with VPI. This patient exhibits hypernasality, characterized by nasalized vowel sounds.  Both the ground truth nasalance and the nasalance estimated by the SI-FT system exhibit three peaks, corresponding to the vowel sounds \textipa{/u:/} and \textipa{/i/} in “Suzie” and \textipa{/i:/} in “sees”. The incomplete VP closure permits excessive nasal airflow and resonance during vowel production, which accounts for the observed hypernasality.

\begin{figure}[htbp]
    \hfill
    \centering
    \vspace{-1mm}
     \includegraphics[width=0.5\textwidth, height=0.32\textheight]{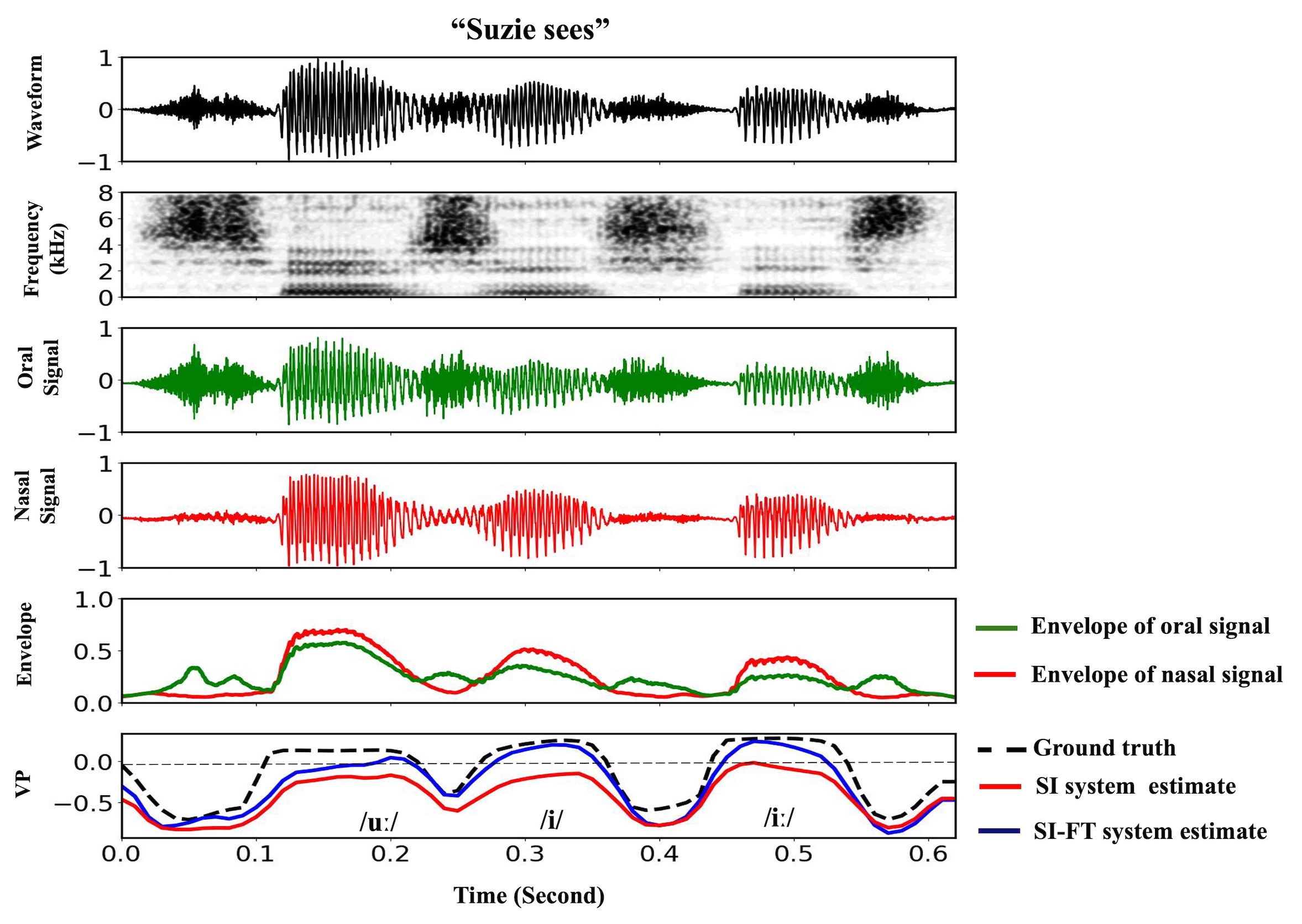}  
    \caption{
    Waveforms, envelope of oral and nasal signals and spectrogram of the utterance “Suzie sees” from a child subject are presented. A comparison between the ground truth nasalance and the corresponding estimates produced by the SI system before and after fine-tuning are shown. The second row displays the spectrogram from the combined signal. The third and fourth rows show the oral and nasal signals, respectively. The fifth row displays the smoothed signal envelopes, with the oral signal shown in green and the nasal signal in red. The sixth row displays the estimated nasalance from the SI and SI-FT systems alongside the ground truth nasalance. 
    }
    \label{fig:nasal_si}
   
\end{figure}

\section{Conclusions And Future Work}
In this study, a novel SI system was proposed for children with VPI, capable of simultaneously estimating nasalance along with three source features. The results demonstrated that the proposed SI system, initially trained on healthy adult speech and subsequently fine-tuned with VPI child speech (SI-FT) significantly improved the estimation of nasalance. This underscores the importance of adapting models to the unique acoustic characteristics of speech in children with VPI. Moreover, the findings highlighted the effectiveness of a multi-task learning framework and the integration of WavLM-Large SSL representations, which collectively enhanced the overall performance of the SI system. The ability of the SI-FT system to estimate nasalance from an audio signal presents a promising non-invasive and cost effective tool for clinical monitoring and assessment of VPI in children.

Future work will focus on developing a SI system using healthy child speech data to better capture the distinct acoustic patterns of children speech.
Additionally, to improve robustness against challenging phenomena such as nasal rustle (the lowest PPMC scores were associated with waveforms exhibiting nasal rustle), the system will be further fine-tuned using a larger collected dataset of VPI-Child speech to include more examples of nasal rustle, thereby enhancing its potential effectiveness in assessing VP dysfunction in children.

\bibliographystyle{IEEEtran}  
\bibliography{IEEEfull}   

\end{document}